\documentstyle[epsf,rotate]{mn}

\newcommand{\eg}{{\sl e.g. }}
\newcommand{\ie}{{\sl i.e. }}

\newcommand{\Msun}{\mbox{M$_{\odot}$}}

\newcommand{\rchis}{\mbox{$\chi^{2}_{\nu}~$}}
\newcommand{\kms}{$\,$km$\,$s$^{-1}$}

\title[Aquila X--1: a low inclination soft X-ray transient]
{Aquila X--1: a low inclination soft X-ray transient}
\author[T. Shahbaz, J.R. Thorstensen, P.A. Charles and N.D. Sherman]{
T. Shahbaz,$^{1}$ J.R. Thorstensen,$^{2}$ P.A. Charles,$^{1}$
and N.D. Sherman$^{2}$\\
$^{1}$Department of Astrophysics, Nuclear Physics Building, Keble Road, 
Oxford, OX1 3RH, UK \\ 
$^{2}$Department of Physics and Astronomy, Dartmouth College, 
6127 Wilder Laboratory, Hanover, NH 03755-3528}

\begin{document}

\maketitle

\begin{abstract}

\noindent
We have obtained $I$-band photometry of the neutron star X-ray transient
Aql X--1 during quiescence. We find a periodicity at 2.487 c~d$^{-1}$,
which we interpret as twice the orbital frequency (19.30$\pm$0.05 h).
Folding the data on the orbital period, we model the light curve
variations as the ellipsoidal modulation of the secondary star. We
determine the binary inclination to be 20$^{\circ}$--31$^{\circ}$ (90 per
cent confidence) and also 95 per cent upper limits to the radial velocity
semi-amplitude and rotational broadening of the secondary star to be
117\kms and 50\kms respectively.

\end{abstract}

\begin{keywords}
binaries: close -- stars: individual: Aquila X--1 -- stars: neutron 
-- X-rays: stars.
\end{keywords}

\section{Introduction}

Aquila X--1(=V1333 Aquilae) is a soft X-ray transient source that shows
type I X-ray bursts (Koyama et al. 1981; Czerny, Czerny \& Grindlay
1987), thereby indicating that the compact object is a neutron star. From
quiescent observations the companion star has been identified to be a
V=19.2 K1$\sc iv$ star (see Shahbaz, Casares \& Charles and references
within). Aql X--1 is known to undergo regular X-ray and optical outbursts
on a timescale of $\sim$1 year (Kaluzienski et al. 1977; Priedhorsky \&
Terrell 1984; Charles et al. 1980) much more frequently than the other
neutron star transient Cen X--4 (McClintock \& Remillard 1990).

Attempts to find the orbital period have revealed many modulations. Watson
(1976) reported an unconfirmed 1.3 day X-ray periodicity during the 1975
outburst. Chevalier \& Ilovaisky (1991) have obtained an 18.97 hr
periodicity from optical photometry during its active state, which they
interpret as being the orbital period.

Recently the RXTE All Sky Monitor showed Aql X--1 to have undergone an
X-ray outburst between late January and early March 1997 (Levine \& Thomas
1997). Ilovaisky \& Chevalier (1997) reported that Aql X--1 was optically in
quiescence by 30 March 1997. In this letter we report on our $I$-band
photometry of Aql X--1 obtained in June 1997, when the source was in
quiescence. We are able to confirm the periodicity detected by Chevalier
\& Ilovaisky (1991) and determine the binary inclination.

\section{Observations and Data Reduction}

We monitored Aql X--1 on the nights of 1997 June 20, 23, 24, 25, and 26
UT, using the 1.3 m McGraw-Hill telescope at Michigan-Dartmouth-MIT (MDM)
Observatory and a direct CCD camera equipped with a $1024^2$ thinned
Tektronix chip which yielded 0''.51 pixel$^{-1}$ at $f/7.5$.  The nights
of June 21 and 22 were clear, but the source was too close to the bright
moon to observe. To suppress scattered moonlight and to obtain maximum
sensitivity to the cool secondary star, an interference filter
approximating the Kron-Cousins $I$-band was used for all the exposures,
which were generally 480 s long.  Except for a very few thin clouds on
June 20, conditions were entirely photometric.  The seeing was variable
but generally usable (FWHM $< 2\arcsec$).  Occasional focus changes and
tracking problems spoiled a few frames. In order to obtain maximum
leverage in the time series analysis we observed Aql X--1 for as long as
we could each night.

For flatfield division we used median-filtered exposures of the twilight
sky taken the same nights as the data.  The flatfield pictures agreed
very well from night to night, except for occasional $\sim 1$ per cent
changes in the shadows cast by dust particles on the CCD window.  Bias
subtraction was accomplished using overscan regions of the picture.  For
all reductions we used IRAF\footnote{IRAF, the Image Reduction and
Analysis Facility, is produced by the National Optical Astronomy
Observatories.} routines.  We measured eight stars (Table 1) in each
picture using the {\tt apphot} package; of a variety of software
apertures, a $1.\arcsec4$ radius gave the lowest scatter among the {\it
differential} magnitudes, the averages of which are listed in Table 1.
Photometric standardization was not attempted. The coordinates in Table 1
are derived from a fit of eight stars from the HST Guide Star Catalog
v1.2 in a single good-seeing frame, and are estimated accurate to $<
0\arcsec.3$ from the scatter of the fit.  The mean differential $I$
magnitudes were computed by averaging over all frames and
rejecting points more than three standard deviations from the median. For
the local standard ($\sim 8\arcsec$ north of Aql X--1), only the standard
deviation of the instrumental magnitudes is given.

\begin{table}
\caption{Differential magnitude and position of local field stars.}
\begin{center}
\begin{tabular}{cccl}
         RA    & Dec         &   $\Delta I$         &  Comments \\ 
      (J2000)  & (J2000)     &   (mags)             & \\ \\
  19 11 16.06 & +0 35 05.6 &  2.272$\pm$0.022     & Aql X--1  \\
  19 11 16.11 & +0 35 13.9 &  ($\pm$0.108)        & local standard \\
  19 11 15.61 & +0 35 01.0 &  2.002$\pm$0.018     & comparison star \\
  19 11 13.99 & +0 34 39.9 & -0.524$\pm$0.009     & \\
  19 11 17.03 & +0 35 30.5 &  0.752$\pm$0.009     & \\
  19 11 17.00 & +0 35 22.0 &  2.576$\pm$0.025     & \\
  19 11 18.63 & +0 35 24.0 &  3.024$\pm$0.038     & \\
  19 11 16.83 & +0 34 55.9 &  2.697$\pm$0.040     & \\ \\
\end{tabular}
\end{center}
\end{table}

\section{The search for the orbital period}

We expect the quiescent optical modulation in Aql X--1 to arise primarily
from the ellipsoidal variations of the secondary star, as it does in
other SXTs. These variations arise since the observer sees differing
aspects of the gravitationally distorted star as it orbits the compact
object (van Paradijs \& McClintock 1995). In theory the modulation should
have two maxima and minima. The two minima may be unequal depending on
the binary inclination, but the maxima should be equal. However, in
practice, the light from the accretion disc (especially in the optical)
contaminates the ellipsoidal variations of the secondary star, making
detailed interpretations of the optical light curves difficult (see
Shahbaz, Naylor \& Charles 1993).

Therefore, we first analysed the optical light curve of Aql X--1 using
the phase dispersion minimization algorithm (Stellingwerf 1975). This
technique is insensitive to the shape of the modulation, but does not
remove the effects of the window function. The method groups the data in
phase bins and seeks to minimize the dispersion within the bins. The
deepest minimum of the statistic is the best estimate of the period. The
PDM spectrum was computed in the frequency range 0.1 to 4.0 c~d$^{-1}$ at
a resolution of 0.01c~d$^{-1}$ with 20 phase bins.

The common method of computing a discrete Fourier transform (DFT) and
halving the estimate of the period is equivalent to assuming that the
maxima and/or minima are of equal depths. This may not be the case, as
the observed ellipsoidal modulation may contain unequal maxima and/or
minima depending on the binary inclination and the contamination by the
accretion disc. A Lomb-Scargle periodogram (Lomb 1976; Scargle 1982) of
the data set was then computed with the same resolution and frequency
range as was used for the PDM periodogram.

Figure 1 shows the PDM and Lomb-Scargle periodgrams, with three
frequencies present at 0.524 c~d$^{-1}$, 1.509 c~d$^{-1}$ and 2.487
c~d$^{-1}$. The suggested orbital period of Chevalier \& Ilovaisky (1991)
is marked at 18.97$\pm0.02$ hr. If this is the orbital period, then it
should appear in the Lomb-Scargle periodogram as a peak at twice the
frequency, \ie at 2.530 c~d$^{-1}$. In both the PDM and Lomb-Scargle
periodograms a periodicity at 2.487 c~d$^{-1}$ in present. Chevalier \&
Ilovaisky's period should also appear in the PDM periodogram, since the
method is insensitive to the shape of the modulation in the light curve;
a periodicity at 1.242 c~d$^{-1}$ is present. We note that no periodicity
at 1.242 c~d$^{-1}$ or 2.487 c~d$^{-1}$ is present in the light curves of
the comparison stars. We therefore assume the peak in the Lomb-Scargle
periodogram at 2.487 c~d$^{-1}$ with a power of 31 to be real.

In Figure 1 we also show the 99 per cent confidence level which allows us
to demonstrate the significance on the peaks detected in the Lomb-Scargle
periodogram. The level was calculated from a Monte Carlo simulation,
which calculates the maximum power in 10~000 sets of Gaussian noise with
mean and variance equal to that of Aql X--1. Random peaks reaching a
power of 7.8 are only found in 1 per cent of the artificial data sets,
hence this defines our 99 per cent confidence level. Thus we conclude
that the power level of the 2.487 c~d$^{-1}$ peak is substantially
greater than 99 per cent level significant. In order to estimate the
uncertainty in the 2.487 c~d$^{-1}$ peak position we created an
artificial data set. with a mean, variance, semi-amplitude, amplitude
difference between the minima and baseline equal to that in Aql X--1. The
observed ellipsoidal modulation is well represented by a sinusoid of
frequency $2f$ which lags 90 degrees in phase, relative to sinusoid at
frequency $f$. The amplitude of the sinusoid at frequency $f$ determines
the difference between the minima in the final light curve. Since this
difference is small (0.01 mags), the ellipsoidal light curve is very
similar to a single sinusoid of frequency $2f$.
Using a Monte Carlo simulation we
recorded the peak power in the Lomb-Scargle periodogram near the peak of
interest, for the artificial data set. This was repeated 10~000 times to
produce good statistics, giving a mean peak at 2.487 $\pm$0.005c~d$^{-1}$
(1-$\sigma$ uncertainty). Since this peak corresponds to twice the
orbital frequency, the orbital period is then 1.244$\pm$0.003 c~d$^{-1}$
(=19.30$\pm$0.05 h).

Chevalier \& Ilovaisky (1991) were taken over a 3 month baseline, when
Aql X--1 was in its active state. The $V$-band modulation they observe is
probably due to a combination of X-ray heating of the secondary star, and
of the accretion disc. Modulations of this kind are probably not
coherent, due to the highly variable nature of the accretion disc, even
more so in the $V$-band (van Paradijs \& McClintock 1995). 
Our observations of Aql X--1 were taken when the source was in quiescence,
and where in the $I$-band the accretion disc contribution is negligible
(Shahbaz, Casares \& Charles 1997). Therefore, what we see in the
$I$-band is the characteristic double humped modulation per orbital
cycle, due to the ellipsoidal shape of the secondary star. We therefore
take the orbital period to be 19.30$\pm$0.05 h (=0.804$\pm$0.002 d).

From Table 1 it can be seen that the scatter in the Aql X--1 data is
similar to stars of comparable brightness. This may suggest that the
modulation we have detected is not real. However, it should be noted that
we can still use the largest amplitude of an ellipsoidal modulation that
could conceivably be hidden within the noise in the data, to constrain
the inclination of the system.

\section{Model Fitting and derived parameters}

Using an orbital period $P$=19.30 h (=0.804 d) we folded the Aql X--1
and comparison star data. As there is no spectroscopic ephemeris defining
phase 0.0 (i.e. superior conjunction of the secondary star) we used an
arbitrary value for T$_{0}$. We then shifted the light in phase so that
the deeper minima corresponded to phase 0.0. [Note that the choice is
arbitrary. In the fitting procedure (see later), one of the free
parameters is a phase shift.] The data were then binned into 15 phase
bins, and the standard error and mean were then calculated for each bin
after discarding the data points which were more than 3-$\sigma$ away
from the mean of each bin. This process removed 4 and 12 discrepant data
points from the Aql X-1 and comparison star data respectively; the total
number of data points was 152. Figure 2 shows the resulting light curves
for Aql X--1 and a comparison star.

We fitted the orbital $I$-band light curve of Aql X--1 with an
ellipsoidal model, similar to that used for the other objects in our
programme (see Shahbaz, Naylor \& Charles 1997 and references within).
The model describes the light curve generated by a Roche lobe
filling star, where each element of area on the surface of the star is
assumed to emit blackbody radiation. The temperature distribution over
its surface is assumed to vary according to Von Zeipel's (1924) gravity
darkening law and a linear limb darkening law is used. For a detailed
description of the model see Shahbaz, Naylor \& Charles (1993). The model
parameters were the binary mass ratio ($q$ =M$_{1}/M_{2}$ where M$_{1}$
and M$_{2}$ are the mass of the compact object and secondary star
respectively), the inclination ($i$) and the effective temperature of the
secondary star (T$_{\rm eff}$). A bright spot, \ie the region where the
accretion stream hits the accretion disc was also included. The free
parameters of the model were a phase shift, the normalisation of the
bright spot and the normalisation of the light curve. Thus the fitting
procedure is not dependent on the choice for T$_{0}$. We performed
least-squares fits to the data using this model, grid searching $q$ in
the range 1--10 and $i$ in the range 5$^{\circ}$--50$^{\circ}$. A T$_{\rm
eff}$ of 4620 K, appropriate for a K1$\sc iv$ star was used. We used a
gravity darkening exponent of 0.05; Sarna (1989) finds the gravity
darkening exponent for convective stars with M$_{2}\leq$0.7 to be 0.05.
The limb darkening coefficient for the $I$-band (9000\AA) and the varying
temperature around the surface of the secondary star was interpolated
from the tables given by Al-Naimiy (1978). The best fit gives a \rchis of
1.45 at $q=3$, $i=25^{\circ}$, with a phase shift of 0.0033$\phi$(=0.027
d). This fit is plotted in Figure 2 as the dashed line. The phase shift
allows us to estimate T$_{0}$ to be HJD 2450623.450$\pm$0.016.

Figure 3 shows the resulting ($i,q$) diagram with the 68 and 90 per cent
confidence regions marked, calculated according to Lampton, Margon \&
Bowyer (1976) for three parameters (the phase shift, bright spot
normalisation and the light curve normalisation), after the error bars
had been rescaled to give a fit with a \rchis of 1. As one can see, the
ellipsoidal variation is only weakly dependent on $q$. The 68 and 90 per
cent confidence regions for the binary inclination are
21$^{\circ}$--30$^{\circ}$ and 20$^{\circ}$--31$^{\circ}$ respectively.

Shahbaz, Casares \& Charles (1997) determine the fraction of light
arising from the accretion disc to be 6$\pm$3 per cent at 6000\AA. With
increasing wavelength the disc contamination decreases (see McClintock \&
Remillard 1986; Shahbaz, et al. 1996 and references within), therefore
the $I$-band disc contamination should be lower. If we assume a 95 per
cent upper limit of 9 per cent for the disc contamination, the effect of
this on the binary inclination is to increase it by at most 1$^{\circ}$.

From Figure 2 it can be seen that there seems to be some systematic trend
in the data of the comparison star. As noted earlier the scatter in the
comparison star data is comparable to that in the Aql X--1 data (see also
Table 1). If we assume that the modulation we observe in the Aql X--1
data is not due to the ellipsoidal modulation of the secondary star, and
what we have fitted is the maximum possible modulation that could be
present, then our fits allow us to determine a 95 percent upper limit of
31$^{\circ}$ to the binary inclination.

\section{Discussion}

\subsection{The period difference}

The quiescent period (19.30 h) we find is slightly different ($\sim$ 2
per cent) from the outburst period (18.97 h) obtained by Chevalier \&
Ilovaisky. Outburst and quiescent period differences are observed in SXTs
(see O'Donoghue \& Charles 1996), but the outburst modulation (caused by the
precession of the accretion disc; Whitehurst 1988) is usually a few per
cent {\it longer} than the orbital period. If we interpret Chevalier
\& Ilovaisky's outburst period as being due to the superhump phenomenon,
then the superhump period is shorter than the orbital period by 2 per
cent, \ie we have a ``negative'' superhump. This may seem unusual, but
negative superhumps have previously been observed in cataclysmic
variables, \eg in V503 Cyg the superhump period is shorter than the orbital
period by 3 per cent (Harvey, Skillman, Patterson \& Ringwald 1995). 

\subsection{Upper limits on $K_{2}$ and $v\sin\,i$}

We can compare Aql X--1 to the other long period neutron star SXT Cen X--4.
Assuming that our inclination is correct, and that both systems have
similar mass neutron stars, then one can estimate $K_{2}$, the
semi-amplitude of the radial velocity curve for Aql X--1 ($K_{2}$ scales
with $\sin\,i$). Using $K_{2}$=146 \kms (McClintock \& Remillard 1990) and
$i=40^{\circ}$ for Cen X--4 and $i<31^{\circ}$ for Aql X--1, we find
$K_{2}<117$\kms (95 per cent upper limit) for Aql X--1

We can also determine the upper limit to the rotational broadening of the
secondary star ($v\sin\,i$) assuming that the secondary star fills its
Roche lobe and its spin is tidally locked to the binary period.
Eliminating $K_{2}$ from the mass function and the equation which relates
the rotational broadening to the radial velocity semi-amplitude and mass
ratio (Wade \& Horne 1986), we obtain the expression

\begin{equation}
v\sin\,i\ = 283 \sin\,i (M_{2}/P)^{1/3}~~~{\rm km~s^{-1}}.
\end{equation}

\noindent
The secondary star must be less massive than a main sequence star of the
same spectral type, 0.8 \Msun. Then using equation (1) with
$P$=19.3 hr and $i<31^{\circ}$ gives $v\sin\,i<50$\kms (95 per cent
upper limit). This value for $v\sin\,i$ is consistent with the limits
obtained by Shahbaz, Casares \& Charles (1997).

\section{Conclusion}

Using $I$-band photometry of Aql X--1 taken during quiescence, we have
detected a periodicity at 2.487 c~d$^{-1}$, which we interpret as twice
the orbital frequency. Folding the data on the orbital period
(19.30$\pm$0.05 hr), we fit the ellipsoidal modulation of the secondary
star. We determine 95 per cent upper limits to the binary inclination,
radial velocity semi-amplitude and rotational broadening of the secondary
star to be 31$^{\circ}$, 117\kms and 50\kms respectively.

\section*{Acknowledgements}

The data analysis was carried out on the Oxford Starlink node using the
$\sc ark$ and Starlink software.

\begin{figure*}
{\epsfxsize=500pt \epsfbox[-100 00 700 750]{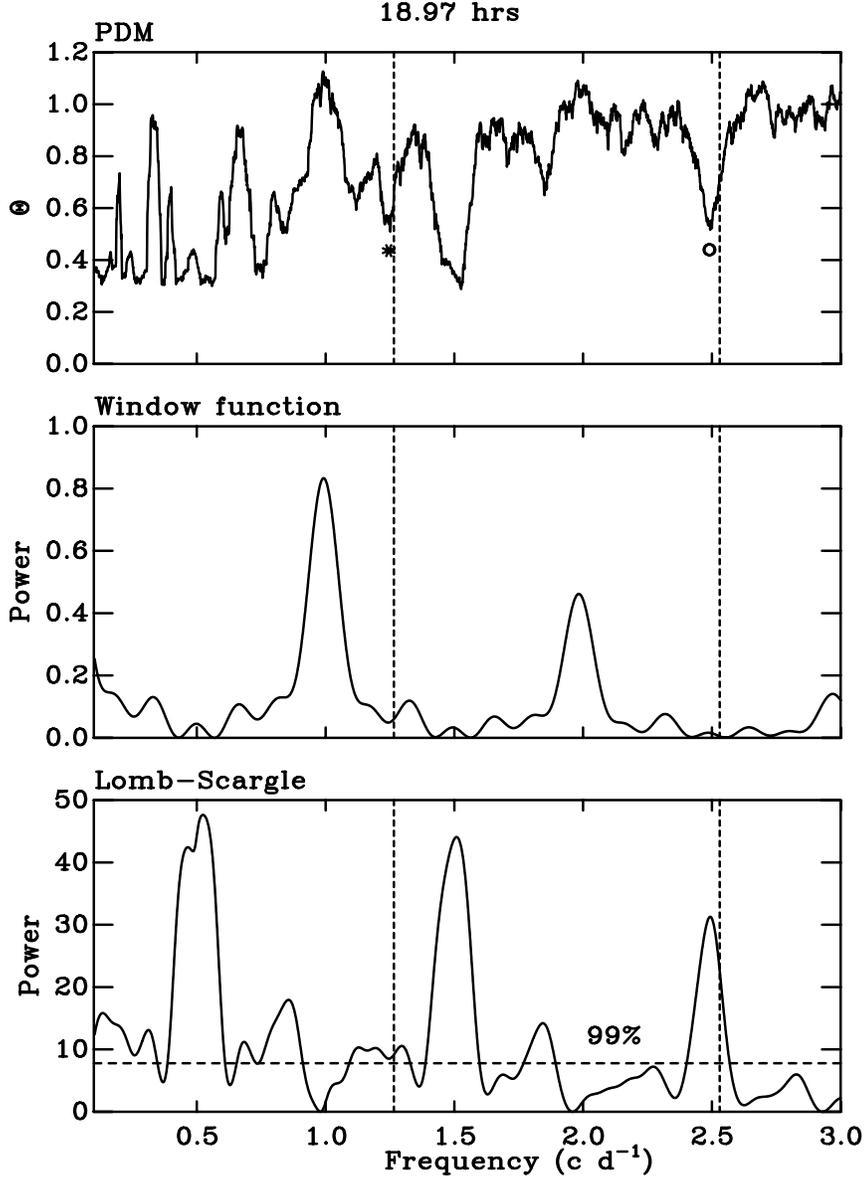}}
\caption{
Results of the period search using the $I$-band data of Aql X--1 during
quiescence. $Top$: Phase Dispersion Minimization (PDM) periodogram. The
frequency 1.242 c~d$^{-1}$ which we interpret as the orbital frequency is
marked with a star; twice this frequency is also shown (white circles).
$Middle:$ Window function. $Bottom:$ Lomb-Scargle power spectrum. The
vertical dashed line is the orbital period determined by Chevalier \&
Ilovaisky (1991) at 18.97 hr. A Monte Carlo simulation provides the 99 per
cent confidence level, shown as the dashed horizontal line. We interpret the
periodicity at 2.487 c~d$^{-1}$, present in the Lomb-Scargle and PDM
periodograms, as twice the orbital frequency (see text), since it is
consistent with twice Chevalier \& Ilovaisky's orbital frequency. }
\end{figure*}

\begin{figure*}
\rotate[l]{\epsfxsize=500pt \epsfbox[00 00 700 750]{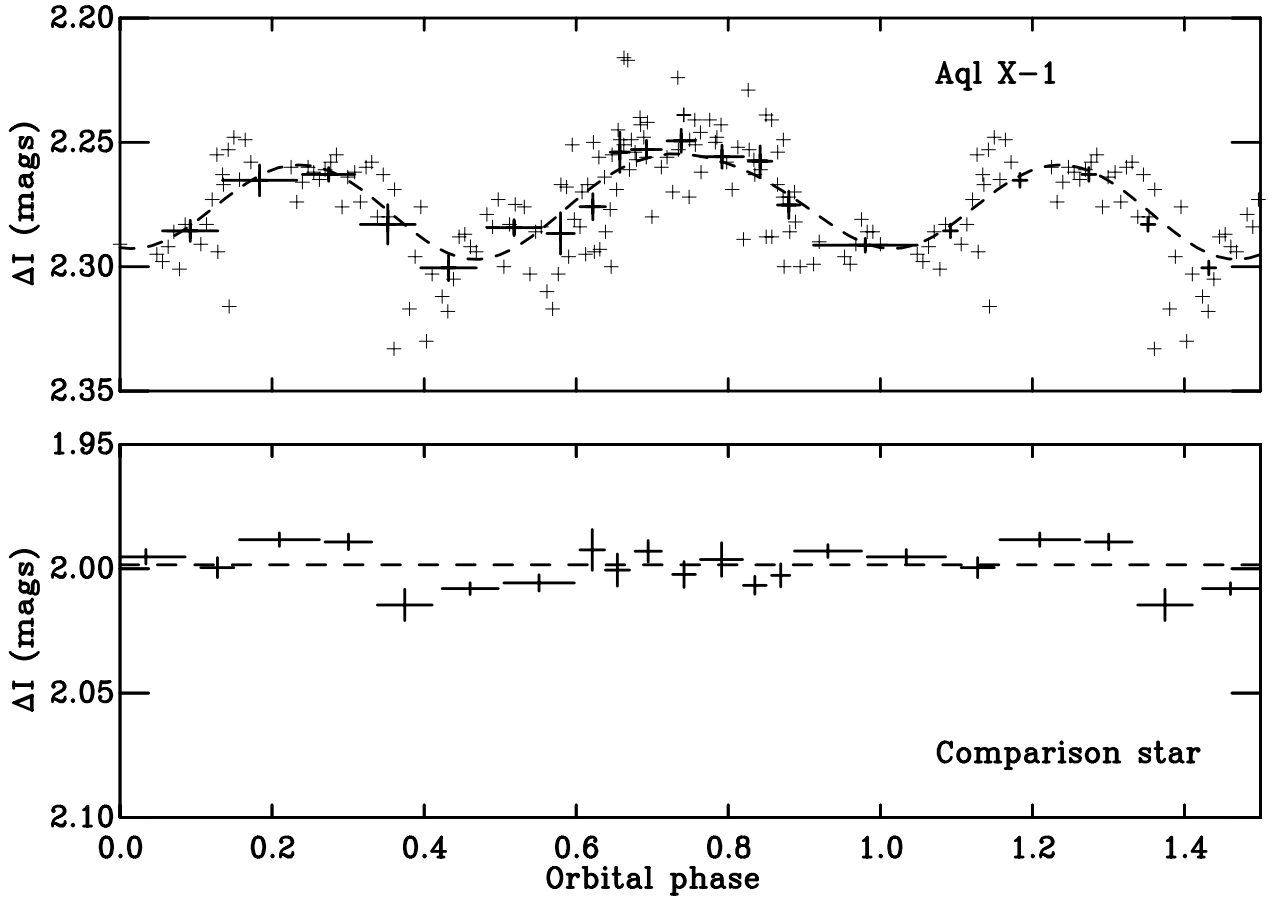}}
\caption{
The phase-binned light curves for Aql X--1 (top) and a comparison star
(bottom). In the top panel we also show the phase-folded data for Aql X-1
(light weighted crosses). The dotted line in the comparison star light
curve is a linear fit. The best fit ellipsoidal model (dashed line) at
$q=3$ and $i=25^{\circ}$ is shown in the light curve of Aql X--1. We show
1.5 orbital cycles for clarity.}
\end{figure*}

\begin{figure*}
\rotate[l]{\epsfxsize=500pt \epsfbox[00 00 700 750]{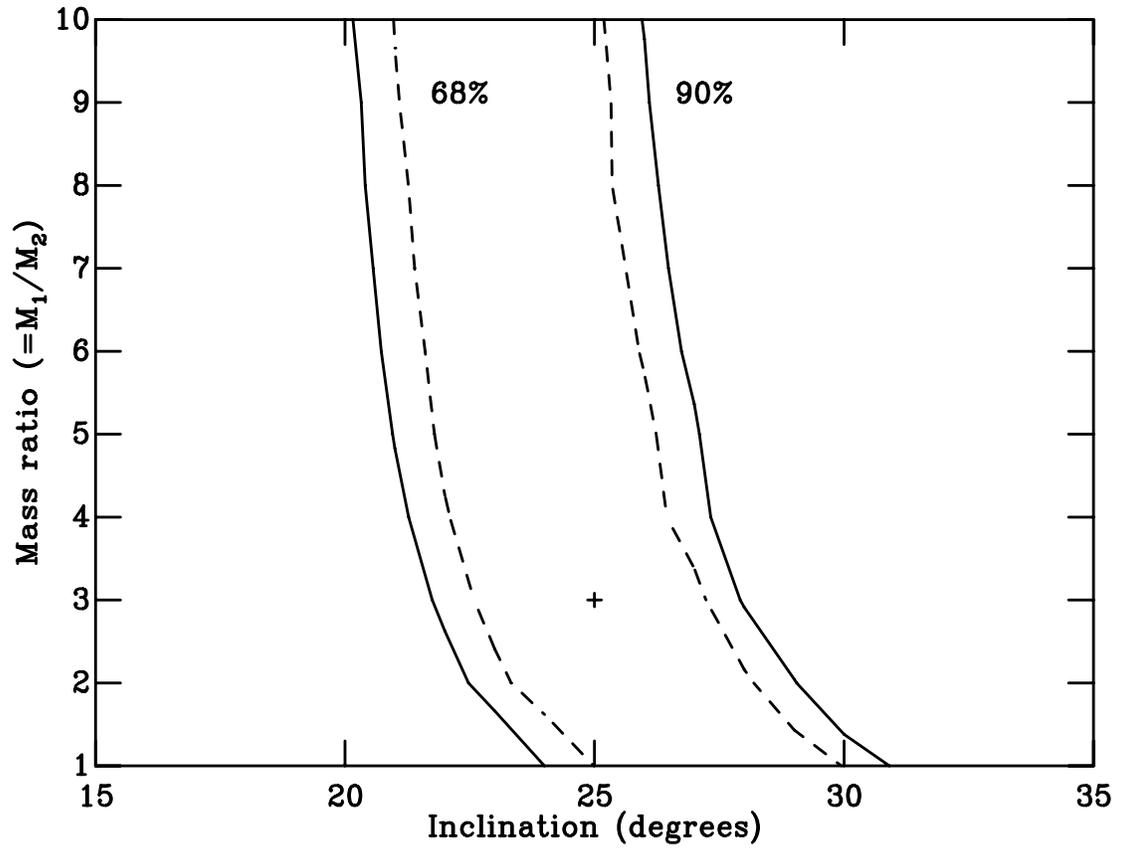}}
\caption{The 68 and 90 per cent confidence level solutions for the
ellipsoidal model fits to the Aql X--1 $I$-band light curve. The cross 
marks the best solution.}
\end{figure*}

\end{document}